\documentstyle[aas2pp4,flushrt]{article}

\begin{document}

\title{On the Self-Consistent Response of Stellar Systems \\
       to Gravitational Shocks}

\author{Oleg Y. Gnedin \& Jeremiah P. Ostriker}

\affil{Princeton University Observatory, Princeton, NJ~08544 \\
       ognedin@astro.princeton.edu, jpo@astro.princeton.edu}

\begin{abstract}

We study the reaction of a globular star cluster to a time-varying tidal
perturbation (gravitational shock) using self-consistent N-body simulations
and address two questions.  First, to what extent is the cluster interior
protected by adiabatic invariants.  Second, how much further energy change
does the postshock evolution of the cluster potential produce and how much
does it affect the dispersion of stellar energies.  We introduce the {\it
adiabatic correction} as ratio of the energy change, $\langle\Delta
E\rangle$, to its value in the impulse approximation.  When the potential is
kept fixed, the numerical results for the adiabatic correction for stars
with orbital frequency $\omega$ can be approximated as $(1 + \omega^2
\tau^2)^{-\gamma}$.  For shocks with the characteristic duration of the
order the half-mass dynamical time of the cluster, $\tau \lesssim
t_{dyn,h}$, the exponent $\gamma = 5/2$.  For more prolonged shocks, $\tau
\gtrsim 4\, t_{dyn,h}$, the adiabatic correction is shallower, $\gamma =
3/2$.  When we allow for self-gravity and potential oscillations which
follow the shock, the energy of stars in the core changes significantly,
while the total energy of the system is conserved.  Paradoxically, the
postshock potential fluctuations {\it reduce} the total amount of energy
dispersion, $\langle\Delta E^2\rangle$.  The effect is small but real and is
due to the postshock energy change being statistically anti-correlated with
the shock induced heating.  These results are to be applied to Fokker-Planck
models of the evolution of globular clusters.

\end{abstract}


\section{Introduction}

The motion of star clusters in the Galaxy is primarily determined by the
monopole component of the Galactic gravitational force; their internal
dynamics is often influenced by the tidal force.  A steady tidal force
imposes a cutoff of the stellar distribution in the cluster at the tidal
radius, whereas fast encounters with the Galactic disk or bulge cause
gravitational shocks (\cite{OSC:72}, hereafter OSC; \cite{S:87}).

The overall effect of tidal shocks is the enhanced evaporation of clusters
via the following processes.  First, stars on average gain energy
($\langle\Delta E\rangle >0$) reducing the cluster binding energy.  After a
short period of contraction immediately following the shock, the cluster as
a whole expands and some stars find themselves outside the tidal boundary.
They are now captured by the galactic potential and are effectively lost
from the cluster.  Second, tidal shocks induce a dispersion of stellar
energies ($\langle\Delta E^2 \rangle >0$), named the ``tidal shock
relaxation'', which is very similar to the diffusion in energy space due to
two-body relaxation (\cite{KO:95}; hereafter KO).  Both of these processes
supply stars to the high-energy tail of the velocity distribution, which is
not bound by the cluster potential.  Those stars eventually escape, thus
enhancing the evaporation of the cluster.  Finally, shock-induced relaxation
speeds up core collapse, which in turn leads to an ever faster evolution,
faster evaporation, and subsequent disintegration of the cluster.

OSC considered the problem in the {\it impulse approximation}, wherein the
stars are assumed to move little during the shock.  This is appropriate in
the outer regions of the cluster, near the tidal radius.  In the cluster
core, the impact of shocks is very different.  The rapidly orbiting stars
smear the perturbation over many rotations around the center of the cluster.
Because the conservation of adiabatic invariants prevents stars from gaining
energy, the shocking effect is significantly reduced.  Spitzer (1987)
studied this problem in the harmonic potential approximation and found that
$\langle\Delta E\rangle$ is exponentially suppressed.  Recently Weinberg
(1994a) suggested that the actual impact of the shock is stronger because of
the numerous resonances in a system with more than one degree of freedom.
Since the original formula for $\langle\Delta E\rangle$ was derived by OSC
in the impulse approximation, we define the {\it adiabatic correction} as
the ratio of the actual energy change to its ``impulsive'' value.

In addition to the remaining question concerning the proper treatment of the
adiabatic correction, there is a much larger open question.  After the
perturbing force has ceased, the cluster finds itself out of equilibrium.
Then there will be an interval of several dynamical times while the cluster
oscillates until phase mixing and Landau damping bring it into a new
equilibrium state.  During this phase, the total energy of the cluster will
be conserved but individual stellar orbits may gain or lose energy in
response to the fluctuating gravitational field.  This may produce
additional both the first and second order energy changes, $\langle\Delta
E\rangle$ and $\langle\Delta E^2\rangle$.  These secondary effects have been
studied very little and may contribute to the cluster evolution, along with
the better known primary effects.

\begin{figure} \plotone{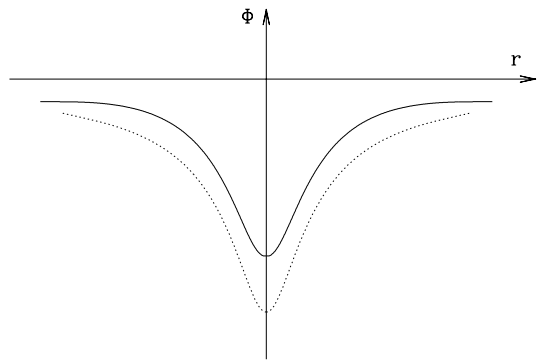}
\caption{Schematic picture of the readjustment of the cluster potential.
  Dots mark the initial King model, solid line represents new equilibrium
  configuration.  (Radial axis is reflected for the illustration purpose.)
  \label{fig:pic}}
\end{figure}

As an instructive exercise, consider how the stars gain energy at the
expense of the binding energy of the cluster in an impulsive shock.  Let
$E_i$ be the energy of the $i$th star,
\begin{equation}
E_i = {1\over 2} m_i v_i^2
      - \sum_{k \neq i} {G m_i m_k \over |{\bf r}_i - {\bf r}_k|},
\end{equation}
where $m_i$, ${\bf r}_i$, and ${\bf v}_i$ are the mass, position, and
velocity of the star, respectively.  The total energy of the system is a sum
of the kinetic ($T$) and potential ($W$) terms:
\begin{equation}
E_{tot} \equiv T + W = \sum_i {1\over 2} m_i v_i^2
      - \sum_{i,k \neq i} {1\over 2} {G m_i m_k \over |{\bf r}_i - {\bf r}_k|}.
\end{equation}
This can be rewritten as
\begin{equation}
\sum_i E_i = E_{tot} + W.
  \label{eq:vir_ei}
\end{equation}
Immediately after the shock, the stars have gained some kinetic energy,
$\Delta E_{i,{\rm sh}}$, but have not yet had time to change the potential.
As the result, the energy of the cluster changes by
\begin{equation}
\Delta E_{tot} = \sum_i \Delta E_{i,{\rm sh}}.
  \label{eq:vir_etot}
\end{equation}
After the period of virialization following the shock, both the kinetic and
the potential energies of the cluster change, $\Delta W_{\rm vir} = -2\,
\Delta T_{\rm vir}$, but the total energy is conserved.  This implies
$\Delta W_{\rm vir} = 2\, \Delta E_{tot}$, and combined with equations
(\ref{eq:vir_ei}) and (\ref{eq:vir_etot}) gives the final energy change of
the stars:
\begin{equation}
\sum_i \Delta E_{i,{\rm vir}} = \Delta E_{tot} + \Delta W_{\rm vir}
  = \sum_i 3\, \Delta E_{i,{\rm sh}}.
  \label{eq:dEnorm}
\end{equation}
The stars gain on average three times the initial kick due to the shock.
The extra energy comes from the reduction of the binding energy of the
cluster.  A schematic picture of the potential change is presented in Figure
\ref{fig:pic}.

We use self-consistent N-body simulations to investigate the adiabatic
correction as a function of the star energy and position in the cluster.
This technique allows us to study directly the self-gravitating response of
the cluster to the tidal perturbation, neglected in prior models.  We find
that collective effects play an important role in transporting the
perturbation from the outer layers deep inside the cluster.  While the
postshock evolution conserves the total energy, the energy of individual
stars changes because of the readjustment of the potential.

A previous approach to include tidal shocks in the evolutionary
Fokker-Planck code (\cite{GO} 1997) has been to first input the energy
changes associated with the shock, and then update the potential using the
adiabatic invariants.  These adiabatic invariants change as a result of
shocking and so the above procedure is not exactly correct, but no better
alternative was available prior to the current work.  Also, this method does
not include the energy transfer between stars subsequent to the shock which
gives rise to additional changes in $\langle\Delta E\rangle$ and
$\langle\Delta E^2\rangle$.  Our current approach is fully self-consistent
and treats all of the above processes exactly.

A direct application of the linear perturbation theory has been done by
Weinberg (1994b,c) and \cite{MW} (1997a-c).  Both methods, our fully
numerical and the semi-analytical by the authors, lead to the correct
results but differ in implementation.

We review the analytic calculations of the adiabatic correction of Spitzer
and Weinberg in \S\ref{sec:adiabatic}.  In \S\ref{sec:code}, we describe the
N-body code and the parameters of the star cluster.  In \S\ref{sec:results},
we present the results of our simulations of disk shocking and compare them
with the analytical estimates.  We consider the problem twice, keeping the
cluster potential fixed, and allowing for self-gravity.  Also in
\S\ref{sec:radial}, we investigate radial shocking to test the effects of
the self-gravitating response of the cluster noted above.  We summarize and
discuss the implications of our results for the Fokker-Planck calculations
of the dynamical evolution of stellar systems in \S\ref{sec:summary}.

% *****************************************************
% *************** ADIABATIC CORRECTIONS ***************
% *****************************************************

\section{Adiabatic Corrections \label{sec:adiabatic}}

We need a parameter to quantify the transition from the impulse to the
adiabatic regime of tidal shocking.  Let $\tau$ be the characteristic
duration of the shock.  We define the {\it adiabatic parameter}
\begin{equation}
x \equiv \omega \, \tau,
\label{eq:xd}
\end{equation}
where $\omega$ is the stellar orbital frequency, or, defined in terms of the
root-mean-square velocity $v_{rms}$ of stars at a distance $r$ from the
center of the cluster, $\omega(r)=v_{rms}(r)/r$.  This frequency is directly
related to the (inverse) dynamical time of stars at a given distance $r$.
The effective time to cross the disk of the Galaxy of half-thickness $H$ for
the cluster with the perpendicular velocity $V$ is
\begin{equation}
\tau = {H \over V}.
\label{eq:tau}
\end{equation}
For the shock duration $\tau$ representative for globular clusters in the
Galaxy, $x \ll 1$ close to the tidal radius, and $x \gtrsim 1$ in the core.
To the second order in perturbation, the initial energy change inside the
cluster relative to the outer parts is independent of the shock amplitude.
Therefore we can define a function $A(x)$ of a single variable $x$ as the
ratio of the actual energy change to its impulse value.  The {\it adiabatic
correction} $A(x)$ should satisfy the following criteria:
\begin{equation}
A(x)=1, \;\; \mbox{for } x\ll 1; \;\;\;\; A(x) \ll 1, \;\; \mbox{for } x \gg 1.
\end{equation}

In the harmonic approximation, the first and second order terms for the
average energy change of stars at the distance $z$ from the equatorial plane
of the cluster are given by Spitzer (1987) and KO, respectively:
\begin{mathletters}
\begin{equation}
\langle \Delta E \rangle = {1 \over 2} \langle (\Delta v_z)^2 \rangle = 
         {2 \, g_m^2 \, \langle z^2 \rangle \over V^2} \, A_1(x),
\label{eq:de_av}
\end{equation}
\begin{equation}
\langle\Delta E^2 \rangle = \langle (v_z \Delta v_z)^2 \rangle = 
         {4 \, g_m^2 \, \langle z^2 \rangle \, v_{rms}^2 \over 3 \, V^2}
         \, A_2(x),
\label{eq:de2_av}
\end{equation}
\end{mathletters}
where $g_m$ is the maximum vertical gravitational acceleration produced by
the disk, and $\langle z^2 \rangle = r^2/3$ for a phase-mixed spherical mass
distribution.  Here
\begin{mathletters}
\begin{equation}
A_1(x) \equiv \exp \left( -{2 x^2} \right),
\label{eq:SpCorr1}
\end{equation}
\begin{equation}
A_2(x) \equiv {9 \over 5} \, \exp \left( -{2 x^2} \right),
\label{eq:SpCorr2}
\end{equation}
\label{eq:SpCorr}
\end{mathletters}
and we denote by the subscripts 1 and 2 the adiabatic correction for the
first and second order terms, respectively.  If we put $A_1=A_2=1$, we
recover the impulse approximation result.  Note that the value of $A_2$
jumps by the factor $9/5$ as $x \rightarrow 0$; i.e., the harmonic
approximation enhances the shock impact at large radii.  However, we should
not expect it to be unity, since the assumption of the parabolic potential
is not valid on the cluster periphery.

The derivation of the above result assumes that the oscillation frequencies
of the system are not commensurable with the perturbation frequency.  In
general, it has been proved that under this condition the adiabatic factor
vanishes faster than any power of $x$ (\cite{K:62}), consistent with the
exponential variation (Spitzer 1987, p. 48).

Recently \cite{W:94} (1994a) showed that resonances do typically occur in a
system with more than one degree of freedom.  If the system is represented
by a combination of multi-dimensional nonlinear oscillators, it becomes
likely that some of the perturbation frequencies will be commensurable with
the low oscillation frequencies of the stars.  Then those stars receive a
significant kick from the perturbation and no longer conserve their actions.
Averaging over an ensemble of stars can give an appreciable energy change.
As a result of summing over the resonant terms, the adiabatic factor $A(x)$
is not exponentially small for the large values of $x$, but rather follows
approximately a power-law form.

\begin{figure} \plotone{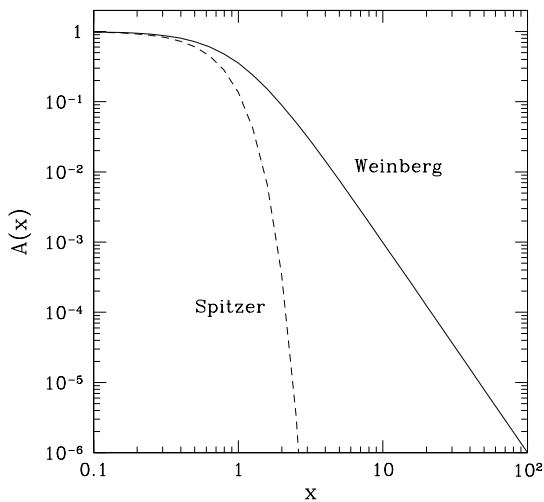}
\caption{Comparison of the Spitzer \mbox{[$A(x)=e^{-2x^2}$]} and Weinberg
  \mbox{[$A(x)=(1+x^2)^{-3/2}$]} adiabatic corrections, i. e., the reduction
  of the energy change due to tidal shocks because of conservation of
  adiabatic invariants of stellar orbits, versus the
  adiabatic parameter, $x$ [eq. (\protect\ref{eq:xd})].  For detailed
  self-consistent simulations, we find that adiabatic corrections show a
  power-law behavior intermediate between these two curves
  (cf. eq. [\protect\ref{eq:SimCorr}] and Table \ref{tab:gamma}).
  \label{fig:AdDisk}}
\end{figure}

Formulae derived by Weinberg (1994b; for example, his eq. [24]) contain
multiple infinite sums and integrals over all frequency range allowed to the
system.  Instead of a direct application of the linear theory formalism, we
derive an asymptotic result for the long duration of the shock, $\tau$, and
fit the intermediate regime.  We follow closely the suggestion by
\cite{T:96}.  Consider a shock with the duration comparable to or larger
than the half-mass radius dynamical time of the cluster.  The acceleration
of stars due to the perturbation is roughly $g_m \, z/H$, and their energies
change significantly only over the period when the stars are in resonance
with the perturbation, $\Delta t \sim \omega^{-1}$.  Thus those stars would
gain about
\begin{equation}
\Delta E_{res} \sim {g_m^2 \, z^2 \over 2 \, H^2 \, \omega^2}.
\end{equation}
The energy change per unit mass is $\Delta E = \Delta E_{res} \times$
(fraction of resonant stars).  Equations (19) and (24) of Weinberg (1994b)
indicate that the number of stars at the peak amplitude scales as $1/\tau$.
Therefore, we approximate the fraction of resonant stars as $(\omega \,
\tau)^{-1}$ and finally obtain
\begin{equation}
\Delta E \sim {g_m^2 \, z^2 \over 2 \, H^2 \, \omega^2} \times 
         {1 \over \omega \, \tau} = {g_m^2 \, z^2 \over 2 V^2}
         \times {1 \over (\omega \, \tau)^3},
\end{equation}
where we have substituted equation (\ref{eq:tau}) for $\tau$.  It follows
from the equation above that the asymptotic form of the adiabatic correction
is
\begin{equation}
A(x) \propto x^{-3}, \hspace{1cm} {\rm for} \hspace{.2cm} x \gg 1.
\label{eq:W1}
\end{equation}
Now we can construct the correction factor for all values of $x$ from its
asymptotic behavior.  The simplest fitting formula is
\begin{equation}
A(x) = \left( 1 + x^2 \right)^{-3/2}.
\label{eq:WCorr}
\end{equation}
In the following, we call equation (\ref{eq:WCorr}) the {\it Weinberg
correction}, as it was derived from considerations brought forward by
M. Weinberg.\footnotemark\
We refer to equation (\ref{eq:SpCorr1}) as the
{\it Spitzer correction}.  We plot the two adiabatic corrections on Figure
\ref{fig:AdDisk}.  The exponential drops much faster than the power-law in
the central region of the cluster.

\footnotetext{ Note that equation (\ref{eq:WCorr}) is an asymptotic fitting
formula for the case $\omega \tau \gg 1$.  The linear theory of Weinberg can
be used (\cite{MW} 1997a-c) in all regimes, albeit the resulting expressions
are rather complex. }

% ************************************
% *************** CODE ***************
% ************************************

\section{Self-Consistent Field Code \label{sec:code}}

We perform N-body simulations of a single tidal shock in a spherical system
in order to investigate the validity of the adiabatic corrections considered
in the previous section.  We use first the self-consistent field (SCF)
method described in \cite{HO} (1992).  The code is designed to reduce the
numerical relaxation, which arises from close two-body interactions.
Instead of direct calculation of forces for all particles, the code computes
orbits of the particles in a smooth potential of the cluster, expanded in a
series of predetermined basis functions.  The coefficients of the expansion
are updated each time step based on the particle positions.  This
procedure provides a self-consistent solution to the Poisson equation.  The
details of the calculation of the expansion coefficients are given in
\cite{HO} (1992), whose set of basis functions we use here.  We have done
simulations with two sets of the expansion coefficients ($n_{max}=6$,
$l_{max}=4$) and ($n_{max}=10$, $l_{max}=6$).  In both cases the results are
indistinguishable and therefore are independent of the numerical method.

As our initial model for the cluster we take the King model (\cite{K:66})
with the structural parameter $W_0=4$ (corresponding to the concentration $c
= 0.84$).  This is a relatively loosely bound cluster, for which the effects
of tidal shocks should be important.  Working units of the code are such
that $G=M=R_c=1$, where $M$ is the total mass of the cluster and $R_c$ is
the core radius.  The tidal radius is determined by the concentration; $R_t
\approx 6.92\, R_c$.  The characteristic time for the whole cluster is the
dynamical time at the half-mass radius $R_h$ (\cite{BT:87}, eq. [2-30]):
\begin{equation}
t_{dyn,h} = \left( {\pi^2 \, R_h^3 \over 2 \, G \, M} \right)^{1/2} \approx 4.5
\end{equation}
in the code units.  For our cluster, $R_h \approx 1.58 R_c$.  We have chosen
a time step which is 1\% of the half-mass dynamical time, $\Delta t = 0.01
\, t_{dyn,h}$.  This time step is sufficient to calculate accurately stellar
orbits even at the center, since the shortest orbital period is $P_c \equiv
4 \, t_{dyn,c} = 4 \, (\rho_h/\rho_c)^{1/2} \, t_{dyn,h} \approx 1.26 \,
t_{dyn,h}$, where $\rho_c$ is the central density, and $\rho_h$ is the
density at half-mass radius; $\rho_h \approx 0.1 \rho_c$ for this cluster.

We use $10^6$ equal-mass particles to model the cluster.  The number of
stars in observed globular clusters is of the same order, so that effects of
``noise'' (such as the Poisson noise) are equally present in both our
simulations and real clusters.  The numerical accuracy of the calculations
is secured by the conservation of the total energy of the cluster in
isolation at the level $\Delta E/E \sim 10^{-5}$.

% *************************************************
% *************** NUMERICAL RESULTS ***************
% *************************************************

\section{Numerical Results \label{sec:results}}

We investigate first the impulsive shocking that occurs during one time
step.  It allows us to study the self-gravitating response of the cluster
separately from the adiabatic corrections.  In run A the potential is fixed,
and in run B it is fully self-consistent.

Then we allow for the time-varying perturbation with the duration of the
order of the dynamical time of the cluster.  For the real clusters in the
Galaxy, a typical shock lasts $\tau \approx 1-3 t_{dyn,h}$.  Run C explores
the case with the fixed potential, run D is the final self-consistent
simulation.  The parameters of all runs are summarized in Table
\ref{tab:runs}.

\subsection{Impulsive perturbation}

Runs A and B study disk shocking in the impulsive regime.  We apply an
impulse of the form
\begin{equation}
{dv_z \over dt} = - I_{imp} \, z
\label{eq:delzshock}
\end{equation}
during one time step at the time $t=t_{dyn,h}$ and then follow the evolution
of the cluster until $t=20\, t_{dyn,h}$.  We take $I_{imp}=1$ in the code
units, which corresponds to $I_{imp} \approx 0.2\, \Delta t^{-1}\,
t_{dyn,h}^{-1}$.  Other components of the stellar velocity remain unchanged
during the impulse.  The overall effect of the shock is characterized by the
relative reduction in the binding energy, $\Delta E_{tot}/E_{bind}=0.002$.

\begin{figure} \plotone{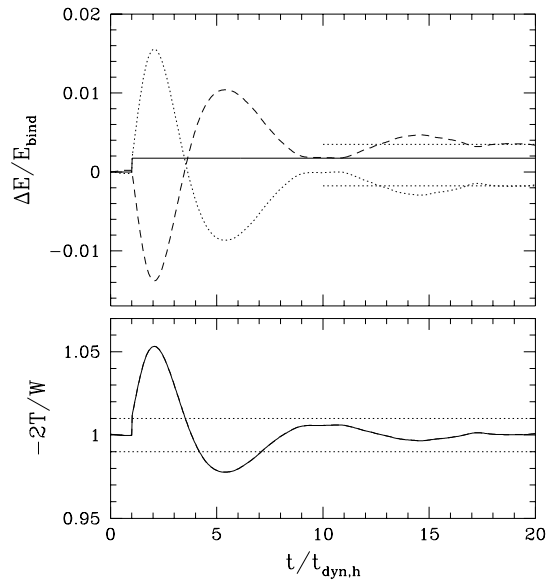}
\caption{Integral characteristics of the cluster after the impulsive
  disk shocking (run B).
  {\it Upper panel}:  the change of the total energy of the cluster
  (solid line) versus time.  Dots and dashes show partial contributions
  of the kinetic and potential energies, respectively.  Energy is
  normalized to the binding energy of the cluster (i.e., the absolute value
  of the central potential), and time is in units of the dynamical time
  at the half-mass radius.  Dotted horizontal lines mark the expected
  values for the kinetic and potential energies after the virialization:
  $\Delta T_{\rm vir} = -\Delta E_{tot}$,
  $\Delta W_{\rm vir} = 2\, \Delta E_{tot}$.
  {\it Lower panel}:  virial ratios $-2\, T/W$ (solid line) and
  $-2\, T/V$ (dashed line) versus time, where $V$ is the Virial
  of the system (eq. [\protect\ref{eq:virial}]).  On this plot both lines
  coincide.  Horizontal dashes mark the 1\% deviation from the equilibrium,
  which we consider as the ``acceptable'' range.  Virial oscillations last
  for about ten dynamical times.
  \label{fig:delz_sg6_e}}
\end{figure}

Figure \ref{fig:delz_sg6_e} illustrates the time evolution of the system.
We plot here the change of the total energy of the cluster, as well as the
partial contributions of the kinetic and potential energies.  The total
energy is very well conserved after the impulse, but both $T$ and $W$
undergo several oscillations.  The lower panel of Figure
\ref{fig:delz_sg6_e} shows deviations from the virial equilibrium as
indicated by the ratios $-2\, T/W$ and $-2\, T/V$, where $V$ is the Virial
of the system
\begin{equation}
V \equiv \sum_k \, {\bf x}_k \cdot \ddot{\bf x}_k,
\label{eq:virial}
\end{equation}
and $V = -2\, T$ for any system of particles moving in a finite region of
space with finite velocities, when averaged over time (e.g., \cite{LL:88}).
Thus, for an equilibrium system the virial ratios should be unity.

From these plots we see how the cluster responses to the perturbation.  At
first, it is compressed since all stars undergo instant acceleration toward
the center.  Their kinetic energy is on average increased and soon they move
outward, and an expansion follows the relatively short period of
contraction.  Now the stars lose kinetic energy, and after about $3\,
t_{dyn,h}$ the virial relations are restored.  However, the cluster
continues to expand following inertia and undergoes a few more oscillations.
The virial equilibrium is approached in about $10-15$ dynamical times.  As
expected, we find $\Delta T_{\rm vir} = - \Delta E_{tot}$ and $\Delta W_{\rm
vir} = 2\, \Delta E_{tot}$ to be within 5\% accuracy at the end of the
simulation.

\begin{figure} \plotone{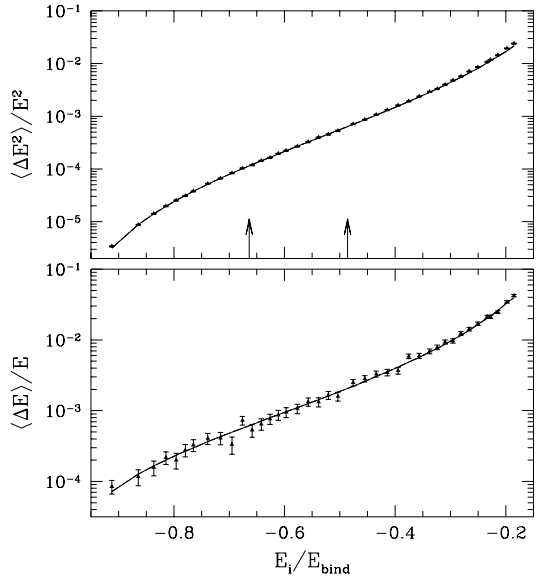} 
\caption{Energy change and dispersion versus initial energy of stars for
  the impulsive shocking in the fixed potential.  Energy along the abscissa
  axis is normalized to the binding energy of the cluster.  Data points
  show the energy changes at the end of the simulation.  The solid lines are
  the analytical predictions (eq. [\protect\ref{eq:dEfix}]).  The center of the
  cluster is on the left; the outer parts are on the right.  Vertical arrows
  indicate the average energy corresponding to the core radius and the
  half-mass radius of the cluster.  The initial tidal radius is near the
  right edge of the plot.
  \label{fig:delz_fix6_b}}
\end{figure}

Now we look at the distribution of the energy changes inside the cluster.
First, we fix the cluster potential (run A) and see how well our results
agree with the analytical predictions.  Figure \ref{fig:delz_fix6_b} shows
the relative changes $\langle\Delta E\rangle/E$ and $\langle\Delta
E^2\rangle/E^2$ at the end of the simulation as a function of the initial
energy, $E_i$.  The stars are grouped into 100 bins, each having $10^4$
particles.  This assures proper averaging over the phase space of stellar
orbits.  We expect that statistical errors do not exceed the 1\% level.  The
solid line shows the expected energy changes for stars of the same initial
energy:
\begin{mathletters}
\begin{eqnarray}
\langle \Delta E \rangle & = & {I_{imp}^2 \, (\Delta t)^2 \over 6}\; r^2 \\
\langle \Delta E^2 \rangle & = & {I_{imp}^2 \, (\Delta t)^2 \over 9}\;
        r^2 \, v_{rms}^2 \, (1+\chi_{r,v}),
\end{eqnarray}
\label{eq:dEfix}
\end{mathletters}
where $\Delta t$ is the time step, $r$ and $v_{rms}$ are the mean position
and rms velocity of stars of energy $E$, and $\chi_{r,v}$ is the
position-velocity correlation function (see \S\ref{sec:app.rv}).  The data
points lie right along the predicted curve, confirming that the binning
procedure is correct and numerical errors are not significant.

\begin{figure} \plotone{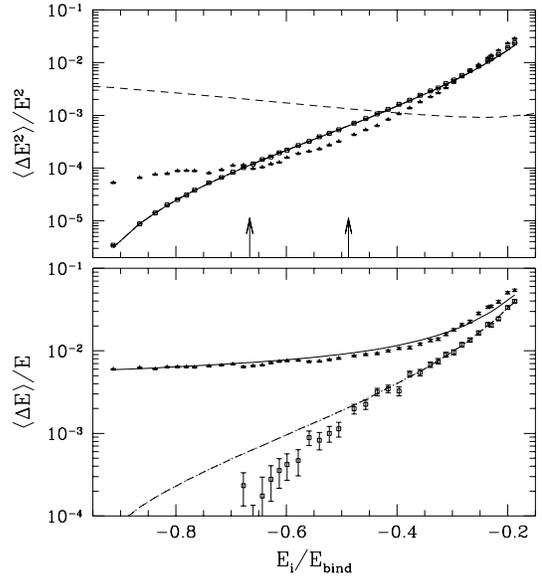}
\caption{Energy change for the impulsive shocking, including self-gravity.
  Open boxes show the results immediately following the shock, solid
  triangle -- at the end of the simulation.  Dashed lines show the
  analytical estimates for the fixed potential (eq. [\protect\ref{eq:dEfix}]).
  $\langle\Delta E\rangle$ is enhanced everywhere in the new equilibrium
  state.  The solid line is our fit (eq. [\protect\ref{eq:dEsg}]).  Dispersion
  is decreased in the middle of the cluster; the apparent flattening in
  the core is due to numerical relaxation (see Text).  The dashes on the
  upper panel show the estimate of the energy dispersion at the end of
  the simulation due to two-body relaxation.\label{fig:delz_sg6_b}}
\end{figure}

Next, we study the self-gravitating response of the cluster to the
perturbation.  Figure \ref{fig:delz_sg6_b} shows the energy changes for run
B.  Right after the shock, both the first and second order energy changes
follow their estimates for the fixed potential.  Twenty dynamical times
later, $\Delta E$ is significantly enhanced everywhere in the cluster.  The
energy dispersion is also much higher in the cluster core.  What causes such
a difference?

First we check whether it is simply a result of numerical relaxation.
Recent studies (\cite{HB:90}; Hernquist \& Ostriker 1992; \cite{W:93})
showed that the self-consistent field method does not totally avoid
relaxation because of the finite number of particles used in calculation of
the potential.  Although reduced, compared to the direct summation methods,
numerical relaxation drives dispersion of stellar energies over the time
scale proportional to the number of particles in the system.  We performed
test simulations varying the number of particles from $N=10^5$ to $N=2
\times 10^6$.  The energy dispersion, $\langle\Delta E^2\rangle_N$, does
decrease with increasing $N$, indicating that there exists a spurious
numerical relaxation.  In the core, the results can be approximated by
\begin{equation}
\langle \Delta E^2 \rangle_N = {const \over N} + 
       \langle \Delta E^2 \rangle_{true}.
\end{equation}

The first order energy change is not affected by the numerical effects as
the dispersion they induce has zero mean.  The observed heating is the
result of the cluster potential attaining a new equilibrium, causing energy
exchange and phase mixing of stellar orbits.

\begin{figure} \plotone{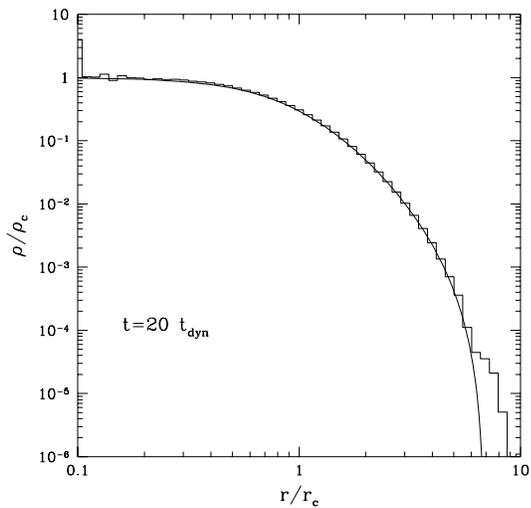}
\caption{Density profile of the cluster at the end of simulation B
  (histogram).  Solid line is the initial King model.  An apparent peak
  in the leftmost bin is an artifact of binning.
  \label{fig:delz_sg6_rho}}
\end{figure}

As a result of the heating, the cluster expands overall.  Figure
\ref{fig:delz_sg6_rho} shows the density distribution at the end of run B.
Some stars move beyond the tidal radius but are not necessarily lost: they
are still gravitationally bound to the cluster.

In the new equilibrium, the depth of the potential well is decreased
relative to the initial configuration (cf. Figure \ref{fig:pic}).  Since the
amplitude of the perturbation is small, the potential changes
self-similarly, i.e. $\Delta \Phi \propto \Phi_i$, where $\Phi_i$ is the
initial potential.  We find that the simulation results can be fitted by an
almost independent sum of the impulsive heating (as calculated by OSC) and
the potential gain:
\begin{equation}
\langle\Delta E\rangle = \langle\Delta E\rangle_{\rm sh} + \Delta E_{pot},
\label{eq:dEsg}
\end{equation}
where $\langle\Delta E\rangle_{\rm sh}$ is given by equation
(\ref{eq:dEfix}), and
\begin{equation}
\Delta E_{pot} = c \; I_{imp}^2 \; (-\Phi_i),
\label{eq:dEpot}
\end{equation}
where $c$ is a normalization constant such that the sum of total energy
change over all particles is twice the total energy change of the cluster
(required by the Virial theorem; cf. eq. [\ref{eq:dEnorm}]).  Note that
$\Delta E_{pot}$ is positive for all stars and is unaffected by the phase
space averaging.

The evolution of the energy dispersion is less obvious.  Figure
\ref{fig:delz_sg6_b} shows that $\langle\Delta E^2\rangle$ {\it decreases}
relative to the impulsive value in the middle of the cluster.  We found this
behavior generic for all simulations we performed.  The distribution of
energies is illustrated in Figure \ref{fig:delz_sg6_d}, which shows the
detailed structure of one bin at the half-mass radius.  The energy change
$\Delta E$ is slightly asymmetric with respect to the center of the bin,
$\langle\Delta E\rangle$, immediately after the shock but it is very well
phase mixed at the end of the run.  The distribution is adequately
represented by the Gaussian form although the dispersion is indeed reduced
by about 20\% relative to the initial impulse.  The normal shape of the
distribution is intrinsic as it is unaffected by varying the size of the
bins.  The skewness and kurtosis of the distribution are small and
statistically insignificant.

The reduction in the dispersion of stellar energies after potential
fluctuations, or the ``cooling'' of the cluster, can be explained by the
anti-correlation of the initial energy change and the subsequent change due
to self-consistent oscillations.  In a compressive shock, stars that were
moving inwards before the shock gain energy and those moving outwards lose
energy.  But, after the shock there is a compressive wave moving inwards
(which subsequently reflects and moves outwards), so those particles which
had gained energy are in phase with the wave and those which had lost energy
are out of phase.  In an expanding shock, all directions are reversed but
the relation between the energy gain and the similarity of phase is
maintained.  This process induces an anti-correlation between the initial
energy change and the subsequent change when the wave reverses.

\begin{figure} \plotone{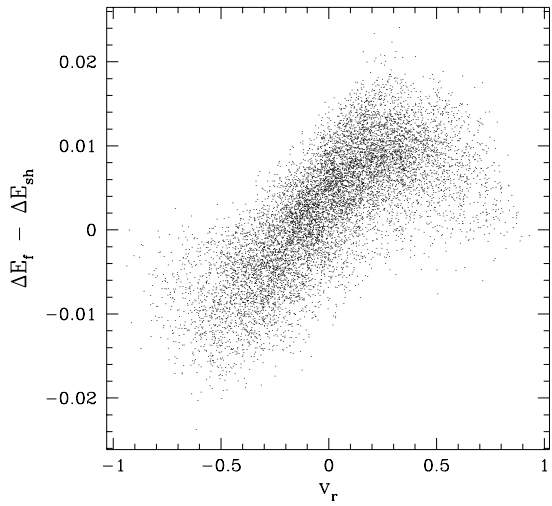}
\caption{Energy change due to potential evolution versus initial
  radial velocity in the bin at the half-mass radius for run B.  Both
  quantities are in the code units.  The postshock energy gain correlates
  with the motion of stars at the moment of shocking.
  \label{fig:devr_delzsg6}}
\end{figure}

\begin{figure} \plotone{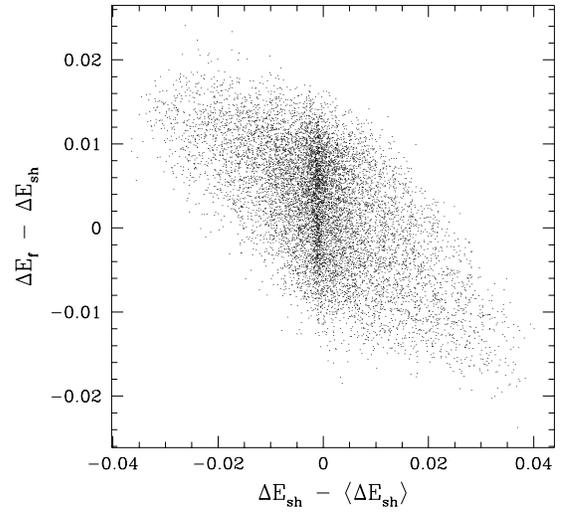}
\caption{Energy change due to potential evolution versus impulsive energy
  kick, $\Delta E_{\rm sh}$, in the bin at the half-mass radius for run B.
  Here $\langle\Delta E_{\rm sh}\rangle$ is the average energy change in the
  bin.  All energies are in the code units.  The anti-correlation of $\Delta
  E_{\rm f} - \Delta E_{\rm sh}$ and $\Delta E_{\rm sh}$ leads to the
  reduction of the dispersion after the virialization.
  \label{fig:dede_delzsg6}}
\end{figure}

Figure \ref{fig:devr_delzsg6} shows the correlation of the additional energy
change following the perturbation ($\Delta E_{pot} \equiv \Delta E_{\rm f} -
\Delta E_{\rm sh}$) with the radial velocity of stars in the half-mass bin.
Figure \ref{fig:dede_delzsg6} demonstrates that this produces an
anti-correlation of $\Delta E_{pot}$ with the initial kick, $\Delta
E_{kin}$.  The phase mixing of stars during potential fluctuations leads to
a very smooth final distribution of energies in the bin (Figure
\ref{fig:delz_sg6_d}).  Since it fits the expected Gaussian function, the
distribution can be described by only the first two moments, $\langle\Delta
E\rangle$ and $\langle\Delta E^2\rangle$.

Another point of view has been suggested by the referee, M. Weinberg (also,
\cite{JHW:98}).  Stellar actions, the radial action $I_r$ and the angular
momentum $J$, change after the impulsive shock.  During a relatively slow
virialization process following the shock, the actions are conserved and so
is the phase space available to stars of the same initial energy.  However,
the energy space has decreased after the virialization, causing the
corresponding decrease of the energy dispersion, $\langle\Delta E^2\rangle$.

We have checked the postshock conservation of the phase space in run B.  The
angular momentum is virtually conserved, although $\langle\Delta J^2\rangle$
increases in the cluster core due to numerical relaxation effects.  The
radial action grows during the virialization, presumably because of the
change of the potential.  The dispersion of the radial action,
$\langle\Delta I_r^2\rangle$, increases in the core due to numerical effects
but decreases slightly in the middle parts of the cluster.  We find that the
value of $\langle\Delta I_r^2\rangle$ drops by 9\% at the half-mass radius
and by 13\% at larger radii.  At the tidal radius, $\langle\Delta
I_r^2\rangle$ again increases relative to the impulsive value.  Although of
lower amplitude than the corresponding reduction of the energy dispersion
(39\% and 34\%, respectively), the {\it decrease} of the dispersion of the
radial action during the postshock phase is not a numerical effect.  This
must be a contribution of the collective oscillatory effects, in addition to
the change of the potential.

% *****************************************

\subsection{Time-varying perturbation}

Having explored the general features of the self-gravitating response of the
cluster to the impulsive perturbation, we now turn to modeling the more
realistic time-varying shocking.  For the disk shocking in the Galaxy, the
vertical gravitational acceleration towards the Galactic plane can be
approximated by a Gaussian function of the height $Z$, with the
characteristic thickness $H$.  If we assume a constant vertical component of
the cluster velocity during the passage through the disk, the tidal force
will vary as $e^{-t^2}$.  Therefore, we apply a shock of the form
\begin{equation}
{dv_z \over dt} = - I_{exp} \, z \, 
   \exp \left( -{(t-t_0)^2 \over \tau^2} \right),
\label{eq:expzshock}
\end{equation}
where $I_{exp}=I_{imp} \, \Delta t /(\sqrt{\pi} \, \tau)$ to give the same
total energy change for $\tau \rightarrow 0$ as in the impulsive shocking.
Here $\tau$ is the effective duration of the shock (eq. [\ref{eq:tau}]),
which we have chosen to be equal to the half-mass dynamical time,
$\tau=t_{dyn,h}$, for runs C and D.  We start the simulation long before the
maximum amplitude of the shock occurs at $t_0 \equiv 4\, t_{dyn,h}$ and then
follow the evolution for $30\, t_{dyn,h}$.

\begin{figure} \plotone{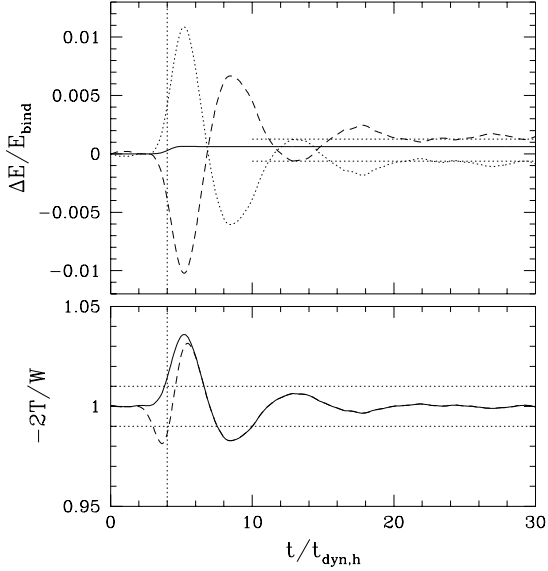}
\caption{Same as Figure \protect\ref{fig:delz_sg6_e}, but for the
  time-varying perturbation (run D).  Vertical dotted line across
  the plot at $t=4t_{dyn,h}$ marks the maximum amplitude of the shock.
  Lower panel shows virial ratios: $-2\, T/W$ (solid line) and
  $-2\, T/V$ (dashed line).
  \label{fig:expz_sg6_e}}
\end{figure}

Figure \ref{fig:expz_sg6_e} shows the total energy change for run D.  This
plot is similar to Figure \ref{fig:delz_sg6_e} for the impulsive shocking,
except that here $\Delta E_{tot}$ is reduced because of the conservation of
the adiabatic invariants of central stars (see also \S\ref{sec:dEtot}).

\begin{figure} \plotone{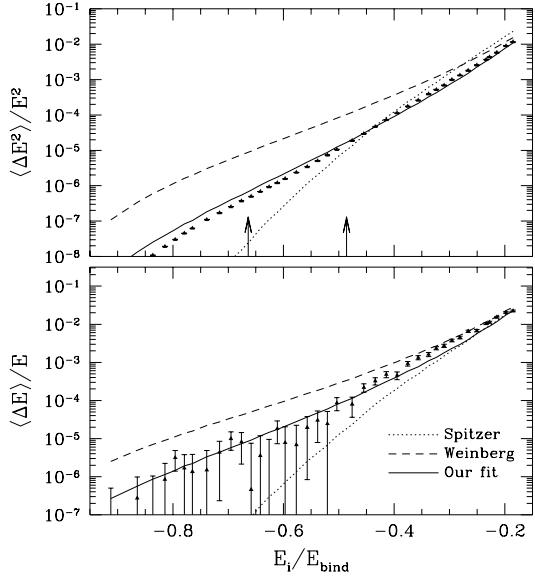}
\caption{Same as Figure \protect\ref{fig:delz_fix6_b}, but for the
  time-varying perturbation in the fixed potential (run C).
  Dots show the analytical expectation with the Spitzer adiabatic
  corrections (eq. [\protect\ref{eq:SpCorr}]), dashes correspond
  to the Weinberg corrections (eq. [\protect\ref{eq:WCorr}]),
  solid lines are our fits (eq. [\protect\ref{eq:SimCorr}]).
  \label{fig:expz_fix6_b}}
\end{figure}

The distribution of the energy changes for the fixed potential (Figure
\ref{fig:expz_fix6_b}) is intermediate of those predicted by Spitzer's
(eq. [\ref{eq:SpCorr}]) and Weinberg's (eq. [\ref{eq:WCorr}]) adiabatic
corrections.  We find that the results in a new equilibrium can be fitted
with a reasonable accuracy by equations (\ref{eq:dEfix}) with the adiabatic
corrections of the form
\begin{mathletters}
\begin{eqnarray}
A_1(x) & = & \left( 1 + x^2 \right)^{-\gamma_1},
\label{eq:SimCorr1} \\
A_2(x) & = & \left( 1 + x^2 \right)^{-\gamma_2},
\label{eq:SimCorr2}
\end{eqnarray}
\label{eq:SimCorr}
\end{mathletters}
with $\gamma_1 = 2.5$ and $\gamma_2 = 3$.

To test the general applicability of these expressions, we have run the
simulations varying the duration of the shock from $\tau = 0.01\, t_{dyn,h}$
to $\tau = 4\, t_{dyn,h}$, and also varying the shock amplitude, $I_{exp}$.
All of the simulation results scale with $I_{exp}^2$, as expected.  We find
that for $\tau \leq t_{dyn,h}$ (on the lower end for the real disk
shocking), equations (\ref{eq:SimCorr}) describe the results of the
simulations very well.  For the more prolongated shocks, the slope of the
adiabatic correction becomes shallower.  Nevertheless, the results can be
fitted by the same functional form, with the exponents given in Table
\ref{tab:gamma}.

Note that for the large values of $\tau$, the overall effect of shocking
decreases ($I_{exp} \propto \tau^{-1}$) and so does the accuracy of the
fits.  However, it is likely that the adiabatic corrections are less steep
for $\tau > t_{dyn,h}$.  In the limit of ``slow shocks'', the asymptotic
Weinberg's result (\ref{eq:WCorr}) becomes valid, as expected.

We have also checked whether the adiabatic corrections are dependent upon
the structure of the cluster.  We have run the self-consistent simulation
for another King model, with the concentration parameter $c = 1.5$.  The
fitting formulae (eq. [\ref{eq:SimCorr}]) slightly underestimate the energy
change and dispersion in the core and the extreme edge of the cluster, but
they hold for most of the intermediate range.  The statistical errors are
small and do not affect the comparison.  Still, it is possible that our
fitting expressions might break for very concentrated clusters and thus
should be taken with caution.

\begin{figure} \plotone{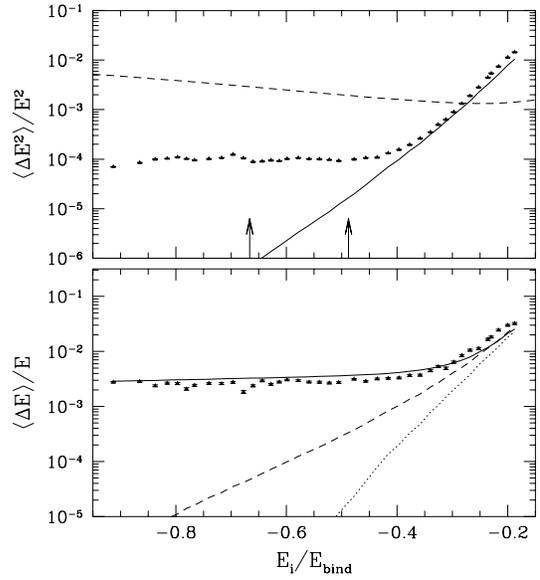}
\caption{Same as Figure \protect\ref{fig:expz_fix6_b}, but for the case
  of the self-gravitating potential (run D).  The solid line in the upper
  panel is the analytic expectation for the fixed potential, the solid line
  in the lower panel is our simple fit including the potential energy
  change (eq. [\protect\ref{eq:dEpot}]).  Dashes on the upper panel show
  the estimate of the energy dispersion at the end of the simulation
  due to two-body relaxation.
  \label{fig:expz_sg6_b}}
\end{figure}

Finally, Figure \ref{fig:expz_sg6_b} shows the energy changes at the end of
run D, the self-gravitating simulation of disk shocking.  As in the
impulsive case, the stellar energies are enhanced everywhere in the cluster.
Our fit to the simulation results has no free parameters: it is a sum of
$\langle\Delta E\rangle$ expected for the fixed potential (eqs.
[\ref{eq:dEfix}] and [\ref{eq:SimCorr}]) and the self-similar potential
change (eq. [\ref{eq:dEpot}]).  Although not exact, the fits are reasonably
good.  Also, we find no additional contribution to the dispersion of energy
due to potential fluctuations.

\subsection{Change of the total energy of the cluster}
\label{sec:dEtot}

One of the predictions of Weinberg's (1994) linear theory is that in the
limit of large $\tau$, the change of the total energy of the cluster due to
a tidal shock with a fixed amplitude is inversely proportional to its
duration.  Since in our simulations the amplitude of the shock scales as
$\tau^{-1}$, to conserve the integral of the perturbing force, we expect the
total energy change to scale as $\Delta E_{tot} \propto \tau^{-3}$.

\begin{figure} \plotone{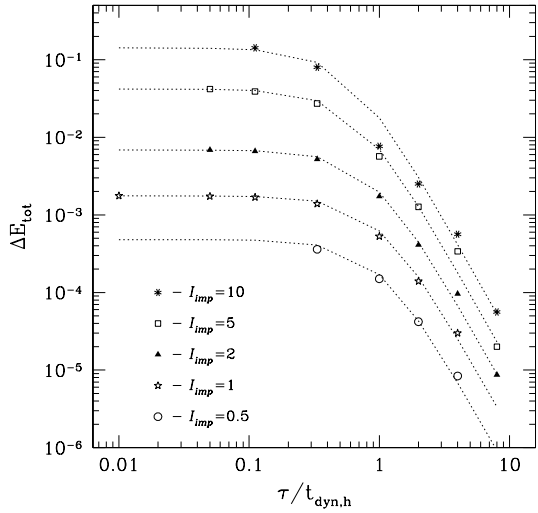}
\caption{Change of the total energy of the cluster due to
  time-varying shocking as a function of its duration for the runs with
  various shock amplitudes.  Dotted lines are our fits
  (eq. [\protect\ref{eq:scaletau}]).
  \label{fig:expz_tau}}
\end{figure}

We use the results of the simulations with various parameters $\tau$ and
$I_{exp}$ to test that prediction.  Figure \ref{fig:expz_tau} shows that in
the limit $\tau \ll t_{dyn,h}$, the value of $\Delta E_{tot}$ is constant.
For the extended shocks, the energy change follows a power-law.  The fitting
formula that combines both limits and provides a good match in the middle
(Figure \ref{fig:expz_tau}) is
\begin{equation}
\Delta E_{tot} = \Delta E_{tot}\vert_{\tau=0} \; \left( 1 + {\tau^2 \over
  t_{dyn,h}^2} \right)^{-3/2}.
\label{eq:scaletau}
\end{equation}
This assures the validity of the linear theory in predicting the integral
characteristics of the system.

% ************************************************************************
% ************************************************************************
% ************************************************************************

\section{Test Case: Radial Shock}
\label{sec:radial}

The main effects of the evolution of the potential, an additional heating in
the core and the decrease of the energy dispersion in the middle of the
cluster, are new.  To check that these effects are not restricted only to
the one-dimensional disk shocking, we investigate another type of tidal
perturbation, {\it radial} shocking, and use a completely independent
numerical code to examine this phenomenon.  The instant acceleration is
proportional to the radius-vector of the star and is directed towards the
center:
\begin{equation}
{d{\bf v} \over dt} = - I_{imp} \, {\bf r}.
\label{eq:delrshock}
\end{equation}
The symmetry of the perturbation in a spherical cluster allows us to use
Spitzer's shell method (\cite{SH:71}) to test the SCF code.  We will see
that the qualitative properties of our solutions survive the change in
numerical methods.

\subsection{The Shell method}

The shell method, originated from an early work by \cite{H:64}, makes use of
the spherical symmetry of the system.  A shell $i$ represents an ensemble of
stars at the radius $r_i$ with the energy $E_i$ and angular momentum $J_i$.
The shell moves in radial direction with the velocity $v_{r,i}$ and can
freely penetrate other shells.  All shells are assumed to have the same
mass, $M_s$.  For shells arranged in order of increasing radius $r_i$, the
equations of motion are
\begin{equation}
{d^2 r_i \over dt^2} = {J_i^2 \over r_i^3} - 
  {(i-\onehalf)G M_s \over r_i^2}.
\label{eq:motion}
\end{equation}
The factor $\onehalf$ takes into account self-gravity of the shell $i$.
While the equations of motion in the spherical system enjoy the simplicity
of Newton's theorem (the force on a shell is determined entirely by the
number of enclosed shells), the calculation of the potential involves all
shells:
\begin{equation}
\Phi_i = -{GM_s i \over r_i} - \sum_{j=i+1}^{N_s} \, {GM_s \over r_j},
\end{equation}
where $N_s$ is the total number of shells.  Therefore, the energy of the
shell, $E_i = v_{r,i}^2/2 + J_i^2/2r^2 - \Phi_i$, depends not only on its
own motion, but also on the position of the shells outside.

This method was successfully applied by Spitzer and collaborators at
Princeton to study the evolution of a spherical star cluster under the
influence of the explicitly imposed perturbations imitating two-body
encounters between stars (see \cite{S:87} for a detailed reference).  A
small energy imbalance resulting from the perturbations was corrected for by
adjusting the kinetic energy of the shells.

Testing the method without external perturbations, we noticed that the total
energy of the system was not conserved.  The energy grew monotonically with
time with the rate proportional to the time step.  The reason for the
nonconservation of energy is the force discontinuity when two shells cross
each other.  Accordingly, we introduced a two-step correction procedure.
First, instead of solving the second order differential equation
(\ref{eq:motion}) as a pair of the first order equations, we integrated the
equation once to obtain a first integral of motion, $C_i$:
\begin{equation}
\left( {dr_i \over dt} \right)^2 + {J_i^2 \over r_i^2} - 
  {(2i-1)GM_s \over r_i} = C_i = {\rm const}.
\label{eq:motion2}
\end{equation}
This equation was then solved numerically to advance shells in radial
direction.  The advantage of solving this equation is in continuity of the
velocity $v_{r,i}$ during shell crossing.  The second correction takes care
of readjusting $C_i$ after each shell crossing.  This procedure assures
conservation of the total energy of the cluster, $E_{tot}$, which has to be
a linear combination of the integrals of motion, $C_i$.  As is easily
checked, $E_{tot} = \sum C_i/2$, with the sum extending over all shells.
Our test runs show exact conservation of $E_{tot}$ in the absence of
perturbations.

Another correction is necessary to solve equation (\ref{eq:motion2}).  When
a shell is close to its turnover points, $r_{max}$ or $r_{min}$, in a given
time step it may jump over to the region of imaginary velocities $v_r$.  The
positions of such shells are integrated in two steps, first up to the
turnover point, and then back with the reversed velocity.

We used the same initial conditions as for the three-dimensional
simulations, projected onto the radial direction.  The angular momentum of
the shells was calculated from their initial positions and velocities and
then kept constant for the rest of the simulation.  We have checked that the
initial conditions represent the assumed King model with correct density
profile and energy distribution.

The nature of the correction procedure for shell crossing is intrinsically
serial, due to the simultaneous adjustment of the constants $C_i$ for both
shells involved in the crossing.  The number of crossings in a given volume
grows very fast with $N_s$, inflating the number of required operations.
The largest run reasonable in computational time involves $10^5$ shells.
This disadvantage is compensated for by the high precision of the results.

\begin{figure} \plotone{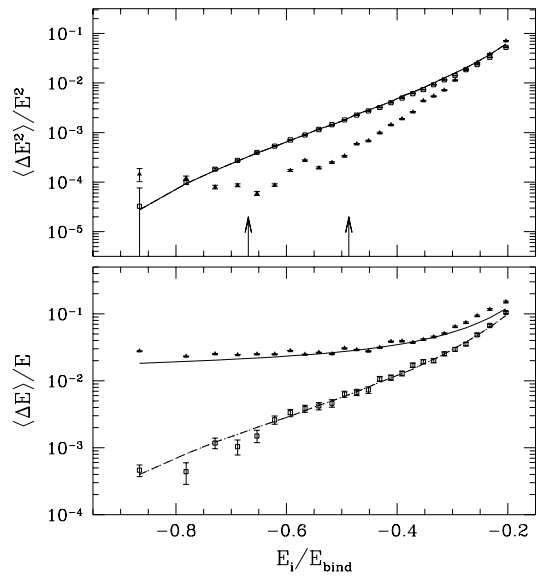}
\caption{Energy change for the impulsive radial shocking using the shell
  method (run E).  Dashes are the analytical estimates for the
  fixed potential, solid line in the lower panel shows the fit
  including the potential energy gain (eq. [\protect\ref{eq:dEshell}]).
  \label{fig:delr_sh5_b}}
\end{figure}

Figure \ref{fig:delr_sh5_b} shows the energy changes for the impulsive
radial shocking (run E).  This plot is qualitatively similar to the
impulsive disk shocking (see Figure \ref{fig:delz_sg6_b}).  We are again
able to fit the results with the sum of the analytical prediction for the
fixed potential and the self-similar potential change:
\begin{mathletters}
\begin{eqnarray}
\langle \Delta E \rangle & = & {I_{imp}^2 \, (\Delta t)^2 \over 2}\; r^2
        + c \; I_{imp}^2 \; (-\Phi_i), \\
\langle \Delta E^2 \rangle & = & {I_{imp}^2 \, (\Delta t)^2 \over 3}\;
        r^2 \, v_{rms}^2 \, (1+\chi_{r,v}).
\end{eqnarray}
\label{eq:dEshell}
\end{mathletters}
This simple fit seems to hold for the radial shocking as well.  However,
once again the energy dispersion is decreased by the postshock evolution.
The effect at the end of the simulation is even stronger than it was in the
case of disk shocking.  This negative feedback must be due to an even
stronger anti-correlation of the effects of the fluctuating potential with
the motion of stars at the time of shocking.

\subsection{Comparison with SCF simulation}

For comparison with the shell method, we have run the three-dimensional SCF
simulation with the radial perturbation (eq. [\ref{eq:delrshock}]).  Run F
involves the same initial conditions as the disk shocking case.

\begin{figure} \plotone{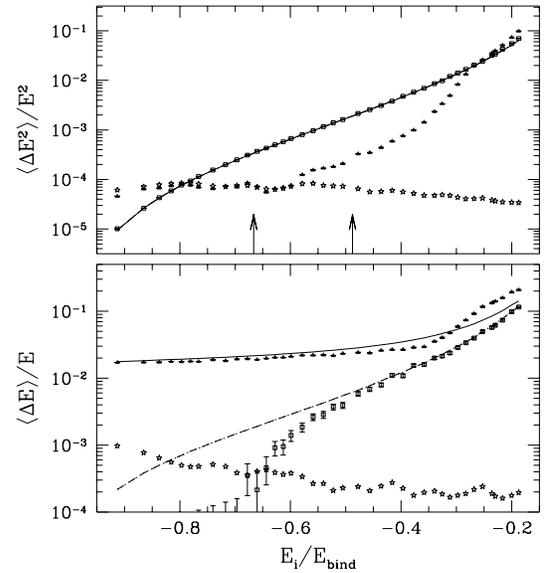}
\caption{Same as Figure \protect\ref{fig:delz_sg6_b}, but for the radial
  impulsive shocking (run F).  The fits are given by equation
  (\protect\ref{eq:dEshell}).  Stars show the numerical errors at the
  end of the simulation as estimated in the isolated run without
  perturbation.  The errors are negligible except for the dispersion
  in the core of the cluster.
  \label{fig:delr_sg6_b}}
\end{figure}

The results of the 3D simulation and the 1D shell method are very similar
(cf. Figures \ref{fig:delr_sh5_b} and \ref{fig:delr_sg6_b}).  The same fit
(eq. [\ref{eq:dEshell}]) is used for both runs and describes $\langle\Delta
E\rangle$ adequately.  The reduction of the dispersion is present in both
simulations, constituting a convincing result.  We have also checked that
numerical errors do not contribute to our results.

% ******************************************
% **************** SUMMARY *****************
% ******************************************

\section{Summary}
\label{sec:summary}

We have explored tidal disk shocking on a star cluster using two independent
types of self-consistent N-body simulations.  As expected, the effects in
the outer parts of the cluster are well described by the classic impulse
approximation, but at the half-mass radius and in the inner parts, the
effects are more complex.

One of the main goals of this paper was to investigate the conservation of
adiabatic invariants of stellar orbits and to obtain convenient adiabatic
corrections that could subsequently be used in the Fokker-Planck
calculations.  Solution of the Fokker-Planck equation is easier and much
faster than full N-body simulations (see, for example, \cite{S:87}).  The
details of our F-P code are given in \cite{GLO:98}.  The Fokker-Planck code
assumes a Gaussian distribution of the energy changes of individual stars
around the mean value, $\langle\Delta E\rangle$, in each energy bin.  We
have checked that the final distribution in our simulations is very nearly
normal (Figure \ref{fig:delz_sg6_d}), allowing us to use the two variables,
$\langle\Delta E\rangle$ and $\langle\Delta E^2\rangle$, to describe the
effects of shocking.  Alternatively, one can use the full formalism of the
linear theory of Weinberg (1994a-c) and \cite{MW} (1997a-c).

First, let us address the mean energy change, $\langle\Delta E\rangle$.  It
is given by equation (\ref{eq:de_av}), where we find the adiabatic
correction $A_1(x)$ given approximately by $(1+x^2)^{-5/2}$ for shock
durations comparable to the half-mass dynamical time, and by
$(1+x^2)^{-3/2}$ for ``slow shocks''.  These results apply when the
postshock evolution of the potential is neglected.  If the latter is
included, the net effect is to produce an additional heating due to the
self-similar potential readjustment, $\Delta E_{pot} = c \, I_{imp}^2 \,
(-\Phi_i)$, where $c$ is a fixed normalization constant (not a free
parameter!).  This second change is qualitatively similar to the method used
by \cite{GO} (1997) for correcting the cluster energies to allow for
adjustment to a new equilibrium, but different in detail in that the exact
treatment allows for the exchange of energy between the stars, which occurs
after the shock.

The second order energy changes have similar adiabatic corrections
(cf. Table \ref{tab:gamma}), varying from $(1+x^2)^{-3}$ to $(1+x^2)^{-7/4}$
as the shock duration increases.  In a self-consistent (time-varying)
potential, the postshock evolution tends paradoxically to {\it decrease} the
energy dispersion because of an anti-correlation between the initial energy
change and the subsequent virialization.

Finally, we found that the change in the total energy of the cluster as a
function of shock duration is in good agreement with the prediction of the
linear theory.  When the integral of the perturbing force is kept fixed,
$\Delta E_{tot} \propto \tau^{-3}$ for $\tau \gg t_{dyn,h}$.

A detailed Fokker-Planck treatment of the cluster evolution will be reported
in \cite{GLO:98}.  But we can note here what changes are expected when the
postshock oscillations are allowed.  We expect that incorporating the
additional first order effect ($\langle\Delta E\rangle$) into the
Fokker-Planck code will increase the rate of cluster evolution, leading to
more rapid core collapse.  Paradoxically, the second order effect
($\langle\Delta E^2 \rangle$) of postshock oscillations appears to lead to a
small reduction in the overall value of the relaxation and a consequent
reduction in the rate of the relaxation-driven evolution.

\acknowledgements

We would like to thank Lyman Spitzer Jr., Jeremy Goodman, Douglas Heggie,
David Spergel and Scott Tremaine for discussions, Antonio Peimbert for
suggesting an efficient implementation of the shell method, and the referee,
Martin Weinberg, for a number of useful comments.  We are especially
indebted to Kathryn Johnston and Lars Hernquist for their help at the
beginning on this project.  Finally, we are grateful to Lars Hernquist for
letting us use his original version of the SCF code.  This project was
supported in part by NSF grant AST 94-24416.

% *************************************************************************
%                                APPENDIX
% *************************************************************************
\begin{appendix}

\section{Parallel implementation of the SCF code}
\label{sec:app.code}

We have implemented the self-consistent field code on two different
platforms, the SGI Power Challenge with shared memory architecture and the
IBM SP2 array with distributed memory.

\begin{figure} \plotone{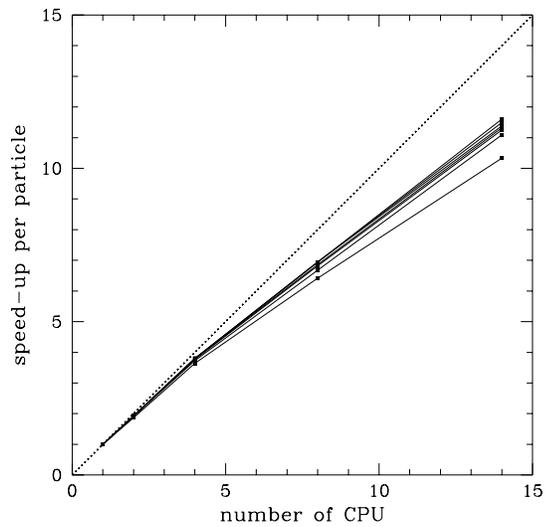}
\caption{Scaling of the computational speed-up per particle per time step
  with the number of CPU on SGI Power Challenge.  The points represent test
  runs with various numbers of particles, from $N=10^4$ to $10^6$.  The
  diagonal dashed line shows the ideal scaling law.  \label{fig:scale_cpu}}
\end{figure}

The benchmark tests on R8000 processors of the Power Challenge show that the
computational time scales perfectly linearly with the number of particles,
from $N=10^4$ to $10^6$.  The speed-up factor per processor is also close to
unity, indicating that the code is ideally parallelizible.  Figure
\ref{fig:scale_cpu} illustrates the speed-up gain per time step per particle
as a function of the number of CPU.  The gain factor is about 3.8, 6.6, and
11.5 for 4, 8, and 14 CPU, respectively.

The performance of the SCF code on SGI Power Challenge can be compared to
its implementation on a Connection Machine 5 reported by \cite{Hetal:95}.
The number of operations per particle depends on the number of coefficients
retained in the expansion of the potential-density pair.  For our test
simulations, we have taken $n_{max}=6$, $l_{max}=4$.  An independent measure
of the performance is the computational time, $t_n$, per time step per
particle per number of the expansion coefficients times the number of CPU.
\cite{Hetal:95} report $t_n \approx 0.9\, \mu$s for their runs (their Table
1).  On the Power Challenge, we get a little better speed, $t_n = 0.65\,
\mu$s.  The average time per time step on a single processor in the run with
$10^6$ particles was 120 s.

We have also done a set of runs on SGI Origin 2000 machine with R10000
processors.  On a single CPU of Origin 2000, the code runs 2.9 times faster
than on the corresponding R8000 processor of the Power Challenge.  However,
the speed-up rate with number of CPU was a little lower.

Finally, we have implemented the SCF code on IBM SP2 at the Maui High
Performance Computing Center using Message Passing Interface (MPI).  Our
preliminary tests show similar performance of the code on SP2 and the Power
Challenge.

\section{The Position -- Velocity Correlation}
\label{sec:app.rv}

We calculate here the correlation function of the mean positions and
velocities of stars of the same energy, $E$.  This function appears in the
analytical expressions for the energy dispersion due to tidal shocks
(eqs. [\ref{eq:dEfix}] and [\ref{eq:dEshell}]).

Consider a spherically-symmetric stellar system.  We seek to establish a
relation of the form
\begin{equation}
{\langle r^2 v^2 \rangle}_E \equiv {\langle r^2 \rangle}_E \,
   {\langle v^2 \rangle}_E \, (1+\chi_{r,v}),
\end{equation}
where the correlation function $\chi_{r,v}(E)$ can depend only on the
energy.  Introducing the integrals
\begin{equation}
I_{n,m} \equiv (4\pi)^2 \, \int\int \, f(E) \, \delta{ \left(
   E-{v^2 \over 2}-\Phi(r) \right)} \, v^{2+n} dv \, r^{2+m} dr,
\end{equation}
where $\delta(x)$ is the delta-function restricting the integration to the
stars of energy $E$, we can write the phase space averaging as
\begin{equation}
{\langle r^2 \rangle}_E = {I_{0,2} \over I_{0,0}}, \;\;\;
{\langle v^2 \rangle}_E = {I_{2,0} \over I_{0,0}}, \;\;\;
{\langle r^2 v^2 \rangle}_E = {I_{2,2} \over I_{0,0}},
\end{equation}
and therefore
\begin{equation}
1+\chi_{r,v} = {I_{2,2} \, I_{0,0} \over I_{2,0} \, I_{0,2}}.
\end{equation}
The integration over the velocity space is straightforward with the aid of
the delta-function.  Also, the distribution function $f(E)$ drops out of
integration since the energy is constant.  Cancelling some constants, we can
redefine the interesting integrals as
\begin{equation}
I_{n,m} \equiv \int_0^{r_m} \, r^{2+m} \, [ E-\Phi(r) ]^{1+n \over 2} \, dr,
\end{equation}
where $r_m(E)$ is the maximum radius a star of energy $E$ can reach,
determined by the following condition: $E \equiv \Phi(r_m)$.  After a simple
dimensionalization of the integrals, we finally define
\begin{equation}
I_{n,m} \equiv \int_0^1 \, x^{2+m} \, \left[ 1 - {\Phi(x r_m) - \Phi_0
    \over \Phi(r_m) - \Phi_0} \right]^{1+n \over 2} \, dx,
\label{eq:corr_int}
\end{equation}
where $\Phi_0$ is an arbitrary constant.  Note that the expression for
$1+\chi_{r,v}$ has an equal number of the $n$ and $m$ indices in the
numerator and the denominator, so that we can cancel from the integration
factors such as $r_m^{3+m}$ and $(2E-2\Phi_0)^{1+n \over 2}$.  Equation
(\ref{eq:corr_int}) can now be solved numerically for any given potential
$\Phi(r)$.  It is useful to choose the value of the constant $\Phi_0$ at the
center of the cluster.

Note, that if the potential obeys a simple power-law, $\Phi(r) = \Phi_0 +
c\, r^{\alpha}$, the integral (\ref{eq:corr_int}) can be done analytically.
The correlation function assumes a particularly simple form,
\begin{equation}
\chi_{r,v} = - {4 \over 10 + 3 \alpha},
\end{equation}
depending only on the power-law index, $\alpha$.  In general, $\chi_{r,v}$
is always negative for self-gravitating systems.  It varies from $\chi_{r,v}
= -0.25$ for the harmonic potential to $\chi_{r,v} \approx -0.57$ for the
Keplerian potential, and lies somewhere inbetween for other systems.

We have calculated the correlation function for a number of King models with
various concentrations, $c$, and fitted the results as a function of the
normalized stellar energy $e \equiv |E/E_{bind}|$:
\begin{equation}
1 + \chi_{r,v} = 0.75 - 0.1 \, (1+c) \, (1-e)
   - 0.1 \, {(1.8 - c) \, c^{1/2} \over 1 + (e/0.15 - 1)^2}.
\end{equation}
The fit is accurate to 2\% for the range of concentration parameters
$c = 0.6$ to $2.7$.

By doing the angular integrals, it is straightforward to show that for
spherical systems with isotropic velocity distribution any combination of
one-dimensional coordinates and velocities has the same correlation
function.  For example,
\begin{equation}
{\langle z^2 v_z^2 \rangle}_E = {\langle z^2 \rangle}_E \,
   {\langle v_z^2 \rangle}_E \, (1+\chi_{r,v}).
\end{equation}

\end{appendix}

% *************************************************************************
%                                 TABLES
% *************************************************************************
\begin{deluxetable}{lccc}
\tablecaption{Parameters of N-body simulations \label{tab:runs}}
\tablecolumns{4}
\tablewidth{420pt}
\tablehead{\colhead{Run} & \colhead{Shocking force} & \colhead{Direction} & 
           \colhead{Cluster potential}}
\startdata
A          &  impulsive     &  vertical  &  fixed            \nl
B          &  impulsive     &  vertical  &  self-consistent  \nl
C          &  time-varying  &  vertical  &  fixed            \nl
D          &  time-varying  &  vertical  &  self-consistent  \nl
E (shells) &  impulsive     &  radial    &  self-consistent  \nl
F          &  impulsive     &  radial    &  self-consistent  \nl
\enddata
\end{deluxetable}

\begin{deluxetable}{lcc}
\tablecaption{Exponents of the adiabatic corrections \label{tab:gamma}}
\tablecolumns{3}
\tablewidth{250pt}
\tablehead{\colhead{Regime} & \colhead{$\gamma_1$} & \colhead{$\gamma_2$}}
\startdata
$\tau \lesssim t_{dyn,h}$  &  2.5  &  3     \nl
$\tau = 2\, t_{dyn,h}$     &  2    &  2.25  \nl
$\tau = 4\, t_{dyn,h}$     &  1.5  &  1.75  \nl
\enddata
\end{deluxetable}

% *************************************************************************
%                              REFERENCES
% *************************************************************************

% *************************************************************************
%                           FIGURE CAPTIONS
% *************************************************************************
%\section{Figure Captions}

%% Figure 1

%% Figure 2

%% Figure 3

%% Figure 4

%% Figure 5

%% Figure 6

%% Figure 7
\begin{figure} \plotone{fig7.ps}
\caption{Distribution of stellar energies in one bin at the half-mass radius
  for run B.  Thin histogram shows the relative energy change immediately
  following the shock, thick histogram shows the final distribution.  Both
  energy changes are normalized to the dispersion at the end of the
  simulation.  Dashed and dotted lines are the Gaussian fits to the
  histograms.  Note (1) the good fit of the final distribution to the Gaussian
  form, and (2) the narrowing (cooling) of the distribution in the final state.
  \label{fig:delz_sg6_d}}
\end{figure}

%% Figure 8

%% Figure 9

%% Figure 10

%% Figure 11

%% Figure 12

%% Figure 13

%% Figure 14

%% Figure 15

%% Figure 16

\end{document}